\documentstyle[prd,aps,preprint,tighten]{revtex}

\begin{document}

\title{How High is High x?}

\author{Jerrold Franklin\footnote{Presented at HIX 2004, Marseille, 
July 26-28, 2004}
\footnote{Internet address:
Jerry.F@TEMPLE.EDU}}
\address{Department of Physics, Temple University, PA 19122-6082}
\maketitle
\begin{abstract}
A general three quark bound state satisfying the Pauli principle, and conserving angular momentum and isospin, is used to investigate the spin structure of  nucleons at high $x$.  It is shown that, if
a spin up quark dominates as $x$$\rightarrow$1
for a spin up proton, it must be a u quark.
Then both $A_{1p}$ and $A_{1n}$, as well as the quark spin distribution $\Delta u/u$, will approach 1 as 
$x$$\rightarrow$1.  The spin distribution $\Delta d/d$ does not approach 1, but is bound by the limits $-\frac{1}{3}\le\Delta d/d\le 0$ 
for any $x$.   The ratio
 \mbox{$F_{1n}/F_{1p}$ will approach $\frac{1}{4}$} (not$\frac{3}{7}$).
\end{abstract}

\newcommand{\up}{\uparrow}
\newcommand{\dn}{\downarrow}
The region of high Bjorken $x$ in deep inelastic scattering (DIS) has become of great interest recently because of the possibility of testing theoretical predictions for the asymmetry parameters $A_{1p}(x)$ and $A_{1n}(x)$
as $x$$\rightarrow$ 1.  A recent experiment has measured $A_{1n}$ up to $x=0.6$\cite{zheng}, and planned experiments may be able to measure the asymmetries at higher x.   Also, there is a renewed recognition that the absence of sea and explicit gluon effects for $x$ above 0.3 makes the high $x$ region a good testing ground for valence quark theories of DIS. 
In this talk, we examine the general properties of nucleon quarks (u and d) obeying the Pauli principle in a general three quark bound state, and show that there are strong model independent constraints relating the behavior of unpolarized and polarized DIS at high $x$.

We start by constructing the most general spin $\frac{1}{2}$, isospin 
$\frac{1}{2}$ bound state of three nucleon quarks.  For a spin up proton, this can be written as\cite{jfpl}\footnote{This state does not include total orbital angular momentum, which we will consider later.}
\begin{equation}
\psi({\rm uud}) = N{\rm uud}\{\up\up\dn[\phi(123)+\phi(213)]
-\up\dn\up\phi(123)-\dn\up\up\phi(213)\}.
\label{eq:gwf}
\end{equation}
This general proton wave function depends on a single three-body spatial function $\phi(123)$, which satisfies the constraint
\begin{equation}
\phi(123)+\phi(213)=\phi(132)+\phi(231).
\label{eq:const}
\end{equation}
The three quark coordinates could be the position, the momentum or, to apply directly to DIS, the Bjorken $x$ of each quark.  The normalization $N$ is model dependent because of interference between $\phi(123)$ and $\phi(213)$.   
The state in Eq.\ (1) is to be considered as the uud part of a completely symmetrized wave function of the form
\begin{equation}
\Psi=\frac{1}{\sqrt{3}}[\psi({\rm uud})+\psi({\rm udu})+\psi({\rm duu})].
\end{equation}
In this symmetrization, the state $\psi({\rm udu})$ is obtained from $\psi({\rm uud})$ by interchanging the second and third spin and space coordinates as well as the indicated quark type, with a similar interchange of the first and third coordinates for $\psi({\rm duu})$.  
For most purposes, including calculating structure functions for DIS,
it is sufficient to use only the uud order\cite{jf1}.\footnote{The uud representation of the proton wave function has been called the ``uds basis" in its application to the Isgur-Karl quark model.  See, for instance, 
Ref.\ \cite{uds} .}   Using all three quark orders would just involve doing the same calculation three times.
 
We arrived at the proton wave function in Eq.\ (1) by the following steps:
\begin{enumerate}
\item We assume that the quarks are antisymmetric in the color degree of freedom, so that the Pauli principle for the two u quarks requires the
spatial and spin functions to appear in the combination 
\mbox{[$\up\dn\up\phi(123)+\dn\up\up\phi(213)$]} in 
the last two terms of Eq.\ (1).
\item For the three quarks to be in a $J$=$\frac{1}{2}$ spin state,  
application of the step-up operator $J_+$ on the state $\psi({\rm uud})$ 
should give zero.  This requires the combination 
\mbox{$\up\up\dn[\psi(123)+\psi(213)]$}
for the first term in Eq.\ (1).
\item The quark-quark forces are charge independent, and the uud state in 
Eq.\ (1) represents a state of isospin $\frac{1}{2}$.  This means that applying the type raising operator d$\rightarrow$u should give zero.  
Doing so gives
\begin{equation}
\psi({\rm uuu}) = N{\rm uuu}\{\up\up\dn[\phi(123)+\phi(213)]
-\up\dn\up\phi(123)-\dn\up\up\phi(213)\}.
\label{eq:uuu}
\end{equation}
We use the fact that the wave function is actually completely symmetrized to rewrite Eq.\ (\ref{eq:uuu}) as
\begin{equation}
\psi({\rm uuu}) = N{\rm uuu}\up\up\dn[\phi(123)+\phi(213)
-\phi(132)-\phi(231)].
\label{eq:uuu2}
\end{equation}
This will equal zero if the constraint of Eq.\ (\ref{eq:const}) holds.
This completes the derivation of Eq.\ (1).
\end{enumerate}
The wave function for a spin down proton is given by the spin lowering operation $\up$$\rightarrow$$\dn$ on Eq.\ (1).  The neutron wave function is given by the type lowering operation u$\rightarrow$d on Eq.\ (1).  This gives,  after rewriting the result in the ddu order,  
\begin{equation}
\psi({\rm ddu}) = -N{\rm ddu}\{\up\up\dn[\phi(123)+\phi(213)]
-\up\dn\up\phi(123)-\dn\up\up\phi(213)\}.
\label{eq:nwf}
\end{equation}

As an example of the completely general nature of the wave function in 
Eq.\ (1), the mixed symmetry wave function $|\!N_M\!\!>$ in the Isgur-Karl model\cite{ik} is given by Eq.\ (1) with
\begin{equation}  
\phi(123)=\Phi^\lambda(2,3,1),
\end{equation}
where $\Phi^\lambda(2,3,1)$ is the Isgur-Karl mixed symmetry spatial function that is symmetric in its first two coordinates. 
If the function $\phi(123)$ in Eq.\ (1) were completely symmetric in all three coordinates, then Eq.\ (1) would represent the simplest form of the wave function for the SU(6) symmetric quark model.   Although SU(6) wave functions have often been given with various linear combinations of the u and d quarks and spin states, they {\em all} can be rewritten in the form of 
Eq.\ (1)\cite{jf1,r,zd}.

In order to use the general three body wave function of Eq.\ (1) for DIS, we integrate the spatial functions over the coordinates of the unstruck spectator quarks.  For the general case, this leads to the quark spin distributions:
\begin{eqnarray}
u\!\!\up\!\!(x,Q^2)&=&2N^2[2\phi({\bf 1}23)^2 +\phi(2{\bf 1}3)^2
+ 2\phi({\bf 1}23)\phi(2{\bf 1}3)]
\label{eq:uup}\\
u\!\!\dn\!\!(x,Q^2)&=&2N^2\phi(1{\bf 2}3)^2
\label{eq:udn}\\
d\!\!\up\!\!(x,Q^2)&=&2N^2\phi(12{\bf 3})^2
\label{eq:dup}\\
d\!\!\dn\!\!(x,Q^2)&=& 2N^2[\phi(12{\bf 3})^2
+\phi(12{\bf 3})\phi(21{\bf 3})].
\label{eq:ddn} 
\end{eqnarray}
In each case, the coordinate in boldface represents the struck quark, and the other two coordinates are integrated over.  The $x$ variable in the quark distribution is that of the struck quark.
The dependence of the quark distributions on $Q^2$, the four-momentum transfer to the nucleon squared, follows from the mechanics of deep inelastic scattering.  We will drop the $Q^2$ variable from the notation, but its presence should be kept in mind.   Our general analysis applies for any $Q^2$ large enough to eliminate interference between two different struck quarks.  Because we are only interested here in the high $x$ region, we do not include sea quarks or explicit gluonic effects which are expected to be small for $x>0.3$. 

Inspection of Eqs.\ (\ref{eq:uup})-(\ref{eq:ddn}) shows that if a spin up quark dominates as $x$$\rightarrow$1, as suggested by perturbative QCD\cite{fj,brodsky},
then the first variable in $\phi(123)$ must dominate.  Dominance of the second or third variable would preserve a spin down quark at large $x$.
Only the spin up u quark  distribution of Eq.\ (\ref{eq:uup}) depends on the first variable in $\phi(123)$, so only that quark can dominate as $x$$\rightarrow$1.  
That is,  if any one quark in a proton dominates as $x$$\rightarrow$1, it must be the u quark with the same spin as the proton.
This means that the ratio 
\mbox{$F_{1n}/F_{1p}$ will approach $\frac{1}{4}$}
as $x$$\rightarrow$1, and the asymmetry functions
 $A_{1p}$ and $A_{1n}$, as well as the quark spin distribution $\Delta u/u$, will approach 1 as $x$$\rightarrow$1.

The quark spin distribution $\Delta d/d$.   
is also of interest as $x$$\rightarrow$1.
This ratio cannot approach 1.
From Eqs.\ (\ref{eq:dup}) and (\ref{eq:ddn}) we see that
\begin{equation}
\frac{\Delta d(x)}{d(x)}=\frac{-\phi(12{\bf 3})\phi(21{\bf 3})}
{[2\phi(12{\bf 3})^2+\phi(12{\bf 3})\phi(21{\bf 3})]}.
\end{equation}
This ratio depends on the interference term in the numerator.  For
SU(6) symmetry, the ratio would be $-\frac{1}{3}$.  It will remain negative if SU(6) breaking is not so radical that the numerator changes sign, which is unlikely since that would require a node in the three quark wave function.  Thus, our model independent prediction is that 
$-\frac{1}{3}$$\le$$\Delta d/d$$\le$0 for all $x$, independently of the behavior of the ratio $F_{1n}/F_{1p}$.
As $x_3$ approaches 1, $x_1$ and $x_2$ will both be small so that 
$\phi(12{\bf 3})$ and $\phi(21{\bf 3})$ will be about equal.
Thus, $\Delta d/d$ is expected to approach $-\frac{1}{3}$ for large $x$. 
Measurements of $\Delta d/d$ up to $x$=0.6 do have this bvehavior \cite{zheng}.

The bound $-\frac{1}{3}$$\le$$\Delta d/d$$\le$0 for all $x$ is model independent, and should hold for any three quark bound state satisfying the Pauli principle, and conservation of angular momentum and isospin.  There has been one prediction\cite{fj,brodsky} that $\Delta d/d$$\rightarrow$1 as $x$$\rightarrow$1, which violates this bound.  A perturbative QCD argument is made in Refs.\ \cite{fj} and \cite{brodsky} that spin up quarks (both the $u\!\!\up$ quark and the $d\!\!\up$ quark) dominate as $x$ becomes large.  While this is possible for the $u\!\!\up$ quark, 
compaison of Eqs.\ (\ref{eq:dup}) and (\ref{eq:ddn}) shows that the $d\!\!\dn$ quark must have comparable strength to the $d\!\!\up$ quark, in any limit.  Thus, the only way for the wave function to be dominated by a spin up quark is if it is only the $u\!\!\up$ quark (and not the $d\!\!\up$ quark) that dominates, as discussed earlier in this paper.  The ratio $\Delta d/d$ is then limited to be between $-\frac{1}{3}$ and 0, and cannot approach 1 for any $x$.  We also see from Eqs.\ (\ref{eq:uup})-(\ref{eq:ddn}) that, if $u\!\!\up$ dominates as $x$$\rightarrow$1, then both $d\!\!\up$ and $d\!\!\dn$ must be negligible as  $x$$\rightarrow$1.  This means that the perturbative QCD assumption that a spin up quark dominates as $x$$\rightarrow$1 should also lead to the prediction that $F_{1n}/F_{1p}$$\rightarrow$$\frac{1}{4}$,
and not $\frac{3}{7}$ which would be the case if a spin up d quark also remained at high $x$.

We have not explicitly included orbital angular momentum,
but we do not expect it to affect our analysis as $x$$\rightarrow$1.  This is because the ratio of perpendicular to longitudinal momentum goes to zero as $x$$\rightarrow$1, leading to $L_z$=0.  The wave function in Eq.\ (1) is general enough that it includes orbital angular momentum if $L_z=0$.  Relativistic effects could reduce $A_1$ because the small components of the Dirac wave function reduce the quark spin projection.  However this effect also depends on the ratio of perpendicular to longitudinal momentum, so that it too does not affect our results for $x$$\rightarrow$1.

The prediction that $A_{1p}$ and $A_{1n}$ each
approach 1  as $x$$\rightarrow$1 
is a challenge for experiment.  Current measurements indicate that $A_{1p}$ may be increasing above the SU(6) value of $\frac{5}{9}$, but it is not clear that the trend around $x=0.6$ is steep enough to reach 1 at $x=1$.  $A_{1n}$ is positive in the new measurement at $x=0.6$\cite{zheng}, 
but is still not close to 1.
Experiments at higher $x$ will be needed to test the $x$$\rightarrow$1 limits.  The experimental tests are complicated by a number of problems.  It is difficult to achieve accurate asymmetry measurements at large $x$.  The $Q^2$ dependence increases (beyond QCD evolution) as $x$$\rightarrow$1.  For $A_{1n}$, unfolding nuclear effects from either deuteron or helium asymmetries becomes more uncertain as $x$$\rightarrow$1.  An important question is how high an $x$ is high enough to test the $x$$\rightarrow$1 predictions.

\end{document}